# E-Monitoring Program Pembangunan Infrastruktur Perdesaan (PPIP) pada Dinas PU.Cipta Karya dan Pengairan Kabupaten Muba

Aidil Afriansyah[1], Leon Andretti Abdillah[2], Ria Andryani[3]

[1,2,3] Program Studi Sistem Informasi, Fakultas Ilmu Komputer, Universitas Bina Darma
Palembang, Indonesia
[1]aidil.afriansyah@gmail.com, [2]leon.abdillah@yahoo.com

**Abstract.** Information technology is widely used as a media monitor various activities. In this study, the authors will utilize IT to monitor the Rural Infrastructure Development Program (PPIP). PPIP is a program of community development at the village level in the framework of the provision of basic infrastructure in rural settlements carried out by the Directorate General of Human Settlements Ministry of Public Works to support the policy of the Indonesian government. PPIP Kab. Muba through the Department of Public Works and Human Settlement Irrigation District Muba as actors in the process of distribution of program development implementation, disbursement, monitoring (monitoring), and reporting. In the implementation process of the realization of all the monitoring data is processed in a conventional manner or format of diverse reports from the field so often goes wrong, the late submission of reports and inaccuracies among reports received by the condition of the field. The system can manage employment targets, reporting a physical realization and financial absorption, process data reporting information on a regular basis, timely, complete and factual as the data obtained directly from implementing supervisory officers in the field. This system is built with web-based information technology using PHP and MySQL database.

**Keywords**: E-Monitoring, Good governance, PPIP.

## 1 Pendahuluan

Perkembangan teknologi informasi (TI) atau komputer telah mengubah cara kerja manusia dalam menjalankan aktifitas sehari-hari. TI telah membawa perubahan yang sangat fundamental bagi organisasi baik swasta maupun publik [1]. Hal ini terjadi karena teknologi komputer mampu berkolaborasi dengan banyak bidang ilmu lainnya, termasuk di bidang pemerintahan [2]. Pada instansi pemerintahan tidak bisa dilepaskan pentingnya teknologi. Teknnologi yang berperan mengumpulkan, mengelola, melaporkan suatu informasi sebagai bahan evaluasi aktifitas kerja.

Monitoring perkembangan proyek merupakan faktor penting keberhasilan suatu proyek. Apabila teknologi diterapkan pada sistem pelaporan (elektronik) dapat dikenal dengan istilah *e-monitoring*. *E-monitoring* merupakan pemantauan dan pelaporan dengan penyampaian data secara elektronik (*online*) serta dapat dipantau secara terus menerus untuk penilaian terhadap kualitas dan efektivitas sistem pengendalian untuk meyakinkan bahwa pengendalian telah berjalan sebagaimana yang diharapkan dan diperbaiki sesuai dengan kebutuhan [3]. Dalam upaya menyelenggarakan pembangunan dan penerapan tatakelola pemerintahan yang baik (*good govermance*) Ditjen Cipta Karya Kementerian PU telah melaksanakan Program Pembangunan Infrastruktur Pedesaan (PPIP). Infrastruktur perdesaan merupakan suatu sistem berupa sarana dan prasarana yang tidak terpisahkan untuk kebutuhan masyarakat desa, seperti menyediakan transportasi, pengairan, drainase dan fisik publik lainnya, yang dibutuhkan untuk memenuhi kebutuhan dasar masyarakat baik sosial maupun kebutuhan ekonomi didesa [4]. Dinas PU Cipta Karya dan Pengairan Kab Muba pada Program PPIP sebagai pelaku pelaksanaan di Kabupaten dan bertanggung jawab untuk membuat laporan dan evaluasi kerja pelaku program PPIP. Pada proses pelaporan selama ini yaitu pejabat pengawas memonitoring langsung setelah proses pencairan tahap I, II dan III. Untuk laporan mingguan tim fasilitator lapangan melaporkan dengan hasil print out atau email tidak hanya itu tentunya dengan kendala-kendala seperti format laporan yang beragam, keterlambatan penyampaian laporan, akses ke desa sasaran cukup jauh dan ketidak faktualnya antara laporan yang diterima dengan kondisi dilapangan. Sehingga permasalahan-permasalahan yang ada sulit untuk ditangani dan evaluasi penilaian kerja program PPIP ikut berpengaruh. Dari permasalahan yang ada menjadi kerangka berfikir dalam membangun system e-monitoring, yang merupakan model konseptual tentang bagaimana teori berhubungan dengan berbagai faktor yang telah diidentifikasi sebagai masalah yang penting [5].

Sistem yang dibangun ini diharapkan dapat mengolah data informasi pelaporan secara teratur, tepat waktu, lengkap dan faktual mengatasi permasalahan-permasalahan yang sering ditemui. Sistem ini memuat pengelolaan target kerja, laporan fisik yang dapat di update langsung pejabat pengawas lapangan dari laporan yang diterima. Memuat pelaporan penyerapan keuangan, penyerapan tenaga kerja yang terlibat, melampirkan foto fisik pembangunan dengan persentase pencapaiannya. Data tersebut akan dikelola kedalam database MySQL. Bahasa pemrograman yang digunakan pada penelitian ini menggunakan pemrograman. PHP.

Sejumlah penelitian yang dijadikan penelitian terdahulu, antara lain: 1) Yuniar [6] melakukan penelitian sebuah sistem yang dibangun dapat memudahkan dalam pencarian data, pencatatan arus kas dan untuk jadwal pengerjaan proyek, jadwal setiap pekerjaan dapat ditentukan sehingga dalam pengerjaannya lebih tepat waktu karena ada pengawasannya dalam aplikasi ini, dan 2) Widyawati [7] Sistem informasi *monitoring* pelaksanaan *service order* membantu dalam penyajian *data service* berikut dengan laporan *progress*-nya yang dapat mengecek *service* mana saja yang mengalami hambatan penyelesaian sehingga dapat dengan segera dicarikan solusinya.

## 2  Metode Penelitian

Objek penelitian ini adalah Dinas PU Cipta Karya dan Pengairan Kab Muba pada Program PPIP sebagai pelaku pelaksanaan di Kabupaten dan bertanggung jawab untuk membuat laporan dan evaluasi kerja pelaku program PPIP. Dalam melakukan monitoring program pembangunan infrastruktur pedesaan yang dilakukan oleh Dinas PU Cipta Karya dan Pengairan masih menggunakan cara konvensional, baik itu monitoring keuangan dan monitoring fisik kegiatan monitoring dimaksudkan merupakan program yang terintegrasi, bagian penting dipraktek manajemen yang baik dan arena itu merupakan bagian integral di manajemen sehari-hari [8]. Sehingga diperlukan sebuah system yang mampu mengelola data kegiatan monitoring PPIP dapat dikenal dengan istilah *e-Monitoring*.

## 2.1 Metode Pengumpulan Data

Dalam perancangan system yang akan dibangun ini, peneliti menggunakan beberapa metode, yaitu: 1) Wawancara kepada pejabat yang terkait untuk kegiatan PPIP di Kabupaten Muba, 2) Studi Pustaka dengan pengumpulan informasi dengan mempelajari buku-buku dan referensi yang berhubungan dengan sistem ini, dan 3) melakukan Observasi yaitu melakukan pengambilan dan pengumpulan data dari menganalisa permasalahan yang ada Dinas Pekerjaan Umum Cipta Karya dan Pengairan Kabupaten Musi Banyuasin untuk kegiatan monitoring PPIP.

## 2.2 Metode Pengembangan Sistem

Metode pengembangan sistem yang digunakan adalah model *waterfall*. Menurut Pressman [9] model *waterfall* adalah model klasik yang bersifat sistematis, berurutan dalam membangun *software*. Fase-fase dalam model *waterfall* adalah: 1) *Communication* merupakan analisis terhadap kebutuhan software, dan tahap untuk mengadakan pengumpulan data yang diperlukan, 2) *Planning* pada tahapan ini akan menghasilkan dokumen *user requirement* atau data yang berhubungan dengan keinginan *user* dalam pembuatan *software*, termasuk rencana yang akan dilakukan, 3) *Modeling* proses ini akan menerjemahkan syarat kebutuhan ke sebuah perancangan *software* yang dapat diperkirakan sebelum dibuat *coding*, 4) *Construction* tahapan inilah yang merupakan tahapan secara nyata dalam mengerjakan suatu *software*, artinya setelah pengkodean selesai maka akan dilakukan *testing* terhadap sistem yang telah dibuat tadi, dan 5) *Deployment* tahapan ini bisa dikatakan final dalam pembuatan sebuah sistem. Setelah melakukan analisis, desain dan pengkodean maka sistem yang sudah jadi akan digunakan oleh *user*.

## 3  Hasil dan Pembahasan

Sistem yang akan dibangun dalam penelitian ini adalah sistem yang dapat diakses melalui jaringan internet berbasis web, dimana sistem *e-monitoring* program pembangunan infrastruktur pedesaan ini dapat digunakan oleh petugas untuk memasukkan data secara langsung di tempat lokasi keberadaan petugas. Sistem *e-monitoring* program pembangunan infrastruktur pedesaan ini dapat memberikan informasi monitoring baik Penyediaan Air Minum dan Sanitasi Berbasis Masyarakat (PAMSIMAS) maupun Pembangunan Infrastruktur Perdesaan (PIP). Selain itu juga dapat memberikan informasi laporan kegiatan dan realisasi target kerja.

### 3.1  Halaman Utama *E-Monitoring*

Halaman utama *e-monitoring* ini pengguna sistem monitoring PPIP ini akan dibagi menjadi tiga hak akses yaitu : 1) admin, 2) petugas, dan 3) kasatker. Dengan hak akses masing-masing berbeda (gambar 1).

Pengguna sistem monitoring program pembangunan infrastruktur pedesaan ini akan dibagi menjadi tiga hak akses yaitu admin, petugas dan kasatker. Pengguna dengan hak akses pertama adalah pengguna admin. Admin pada sistem *e-monitoring* program pembangunan infrastruktur pedesaan ini bertugas melakukan penginputan data desa, data kecamatan, dan data kegiatan. Untuk pengguna dengan hak petugas bertugas melakukan pelaporan monitoring kegiatan. Dan Kasatker (Kepala Satuan Kerja PPIP) dengan hak akses menerima laporan kegiatan, berupa laporan rencana target keuangan, realisasi keuangan, rencana target fisik kegitan dan realisasi fisik kegiatan. Laporan yang diterima sebagai bentuk pengawasan dan penilaian kegiatan PPIP.

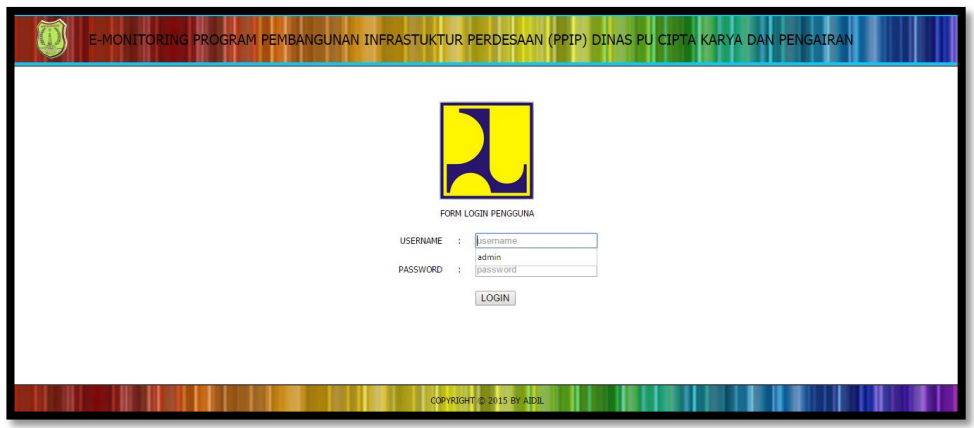

**Gambar 1.** Halaman Utama *E-Monitoring*

### 3.2  Halaman PIP Pencairan Tahap I

Halaman PIP pencairan tahap I adalah halaman yang digunakan oleh petugas untuk memasukkan data hasil monitoring untuk pencairan dana tahap 1 untuk program PIP sampai dengan pencairan tahap 3, baik itu program PIP maupun kegiatan Pamsimas (gambar 2).

**Gambar 2.** Halaman Form Input Pencairan

### 3.3  Halaman Grafik Target Kerja

Halaman grafik target kerja adalah grafik dan tahap pemeriksaan dimana pada masing-masing tahapan tersebut terdapat dua buah grafik batang sebagai pembanding antara rencana dan realisasi (gambar 3).

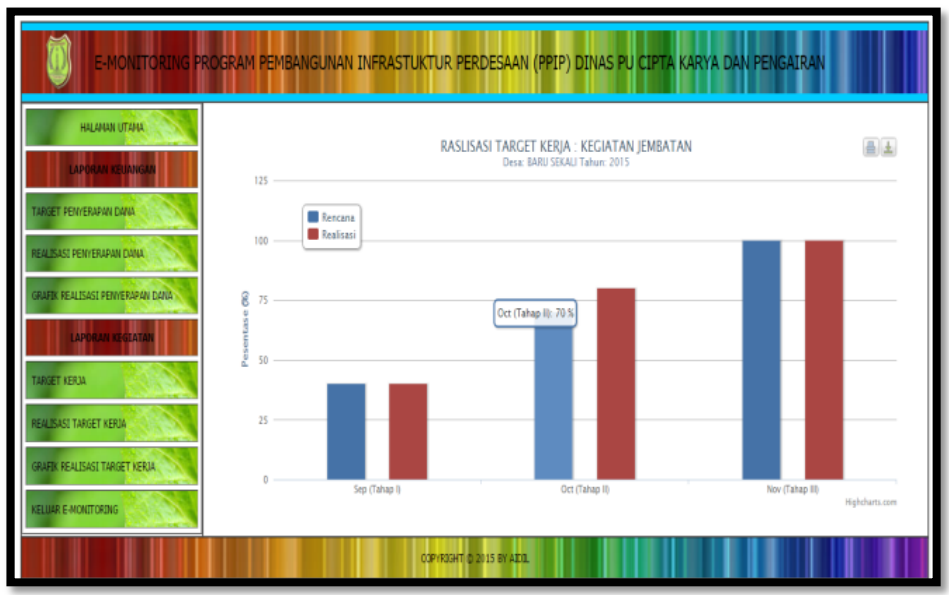

**Gambar 3.** Halaman Grafik Realisasi Target Kerja

# 4  Kesimpulan

Berdasarkan analisis, perancangan, dan konstruksi yang telah diuraikan di atas, maka dapat disimpulkan:

1. Sistem *e-monitoring* program pembangunan infrastuktur perdesaan (PPIP) Dinas PU Cipta Karya dan Pengairan Kabupaten Muba telah dikembangkan dengan bahasa pemrograman PHP dan basis data MySQL.

2. Sistem *e-monitoring* program pembangunan infrastuktur perdesaan (PPIP) Dinas PU Cipta Karya dan Pengairan Kabupaten Muba dapat memproses hasil monitoring pencairan dana tahap 1, tahap 2 dan tahap 3 baik untuk PIP maupun PAMSIMAS. Serta dapat menampilkan laporan kegiatan kecamatan, desa dan realisasi kerja dari target kerja yang ditentukan.

3. Sistem *e-monitoring* PPIP ini telah berjalan sesuai dengen fungsinya hal tersebut ditunjukkan dari hasil pengujian yang menyatakan semua fungsional sistem dapat diterima. Sistem *e-monitoring* program pembangunan infrastuktur perdesaan (PPIP) Dinas PU Cipta Karya dan Pengairan Kabupaten Muba dapat diterapkan dalam melakukan *monitoring*.

# Daftar Pustaka

1. L. A. Abdillah and D. R. Rahardi, "Optimalisasi pemanfaatan teknologi informasi dalam menumbuhkan minat mahasiswa menggunakan sistem informasi," *Jurnal Ilmiah MATRIK*, vol. 9, pp. 195-204, Agustus 2007.

2. L. A. Abdillah, *et al.*, "Pengaruh kompensasi dan teknologi informasi terhadap kinerja dosen (KIDO) tetap pada Universitas Bina Darma," *Jurnal Ilmiah MATRIK*, vol. 9, pp. 1-20, April 2007.

3. N. D. P. P. Mudjahidin, "Rancang Bangun Sistem Informasi Monitoring Perkembangan Proyek Berbasis Web Studi Kasus Di Dinas Bina Marga Dan Pemantusan," *Jurnal Teknik Industri*, vol. 11, pp. 75-83, 2010.

4. Direktorat Jendral Cipta Karya, "Petunjuk Operasional Kegiatan Tahun Anggaran 2014," D. B. Program, Ed., ed. Jakarta: Kementrian Pekerjaan Umum, 2014.

5. Sugiyono, *Metode Penelitian Pendidikan Pendekatan Kuantitatif, Kulaitatif dan R & D. Alfabeta : Bandung*. Bandung: Alfabeta, 2010.

6. I. Yuniar, BK, Raswyshnoe., Nugraha, Nandang, Gita, "Aplikasi Berbasis Web Realisasi Anggaran dan Monitoring Proyek Pekerjaan (Studi Kasus pada PT Dadali Cipta Mandiri)," *Jurnal Prodi Komputerisasi Akutansi Politeknik Telkom Bandung*, 2011.

7. D. K. Widyawati, "Perancangan Sistem Informasi Monitoring Pelaksanaan Servis Order Pada Bagian Perawatan IT (Information Technologi)," *Jurnal Ilmiah Esai*, vol. 6, 2012.

8. D. J. Casely and K. Kumar, *Project monitoring and evaluation in agriculture*. Baltimore: Johns Hopkins University Press, 1987.

9. R. S. Pressman, *Software Engineering: A Practitioner's Approach*, 7th ed. New York, US: McGraw-Hill, 2010.